# Caloric curves for the 8 GeV/c p⁻, ̄p + $^{197}$Au reactions

A. Ruangma, R. Laforest*, E. Martin, E. Ramakrishnan⁺, D. J. Rowland*, M. Veselsky,
E. M. Winchester, S.J. Yennello

*Department of Chemistry and Cyclotron Institute, Texas A & M University,
College Station, Texas 77840*

L. Beaulieu#, W.-c. Hsi^Δ, K. Kwiatkowski§, T. Lefort‡, V.E. Viola

*Department of Chemistry and IUCF, Indiana University Bloomington, IN 47405*

A. Botvina

*GSI, D-64220 Darmstadt, Germany
and Institute for Nuclear Research, 117312 Moscow, Russia*

R. G. Korteling

*Department of Chemistry, Simon Fraser University, Burnaby, BC, V5A 1S6 Canada*

L. Pienkowski

*Heavy Ion Laboratory, Warsaw University, 02 097 Warsaw Poland*

H. Breuer

*Department of Physics, University of Maryland, College Park, MD 20742*

S. Gushue, L. P. Remsberg

*Chemistry Division, Brookhaven National Laboratory, Upton, NY 11973*

B. Back

*Physics Division, Argonne National Laboratory, Argonne, IL 60439*

* Present address: Mallinckrodt Institute of Radiology, St. Louis, MO 63110
⁺ Present address: Microcal Software, Inc., Northampton, MA 01060
# Present address: Department de Physique, Université Laval, Quebec, Canada G1K 7P4
§ Present address: Los Alamos National Laboratory, Los Alamos, NM 87545
‡ Present address: LPC Caen, 6 Boulevard Marechal Juin, 14050 Caen cedex, France
Δ Present address: Rush Presbyterian St. Lukes Medical Center, Chicago, IL 60612

**Abstract**

The relationship between nuclear temperature and excitation energy of hot nuclei formed by 8 GeV/c negative pion and antiproton beams incident on $^{197}$Au has been investigated with the ISiS 4π detector array at the BNL AGS accelerator. The double-isotope-ratio technique was used to calculate the temperature of the hot system. The two thermometers used (p/d-$^{3}$He/$^{4}$He) and (d/t-$^{3}$He/$^{4}$He) are in agreement below E*/A ~ 7

MeV when corrected for secondary decay. Comparison of these caloric curves to those from other experiments shows some differences that may be attributable to instrumentation and analysis procedures. The caloric curves from this experiment are also compared with the predictions from the SMM multifragmentation model.

**I. Introduction**

Investigation of the thermodynamic properties of hot nuclei formed in nucleus-nucleus collisions is one of the major objectives of both theoretical and experimental nuclear science. Recently, one of the most interesting and debated questions in this field has been the possibility of a liquid-gas phase transition in finite nuclei [1-5]. In 1995, the ALADIN collaboration [6,7] presented data suggestive of such a phase transition in peripheral Au + Au collisions at an incident energy of E/A = 600 MeV. The heating, or caloric curve (the relationship between the temperature and the excitation energy of the hot system), was obtained using the double-isotope-ratio temperature technique [8] with excitation energies reconstructed on an event-by-event basis. The results showed a temperature rise with increasing excitation energy per nucleon up to $E^*/A \sim 3$ MeV, followed by a plateau of nearly constant temperature near $T \sim 5$ MeV for excitation energies up to $E^*/A \lesssim 10$ MeV. For higher excitation energies the initial ALADIN results [6] were suggestive of an increase in temperature, analogous to the heating of the free nuclear gas. These studies were presented as evidence for a first-order phase transition, encouraging many research groups to investigate the dependence of nuclear temperature on the excitation energy [9-18].

In contrast, the EoS collaboration data [9-11], obtained for 1 GeV/nucleon Au+C in reverse kinematics, show a monotonic increase of temperature with excitation energy per nucleon with no indication of a plateau. This result suggests a continuous phase transition instead of a first-order phase transition. The caloric curves from ISiS experiment E228, 4.8 GeV $^3$He+$^{nat}$Ag,$^{197}$Au, show an increase in temperature at low E*/A, a distinct slope change near 2-3 MeV, and a gradual increase in temperature up to E*/A = 10 MeV, but no plateau [12,13]. More investigations of caloric curves have been done [14-21]. However, the experimental conditions varied in all of these measurements, complicating the interpretation of the underlying physics, which will be discussed in more detail below. Whether the liquid-gas phase transition of nuclear matter is a first-order or continuous phase transition remains an important question in understanding the decay of hot nuclei.

Hadron- or light-ion-induced multifragmentation of heavy nuclei provides several unique advantages for the study of the nuclear liquid-gas phase transition. Such collisions can create highly excited nuclei via hard nucleon-nucleon scattering, Δ and other hadronic resonance excitations and pion reabsorption [22-24]. Dynamical effects due to collective degrees of freedom, such as rotation, shape distortion, and compression, are minimal in hadron- or light-ion-induced reactions. Therefore, the breakup of the excited system is primarily driven by thermal and Coulomb effects. In this report, the caloric curves for 8 GeV/c $\pi^-$ and $\bar{p}$ beams incident on a $^{197}$Au target have been investigated.

**II. Experimental details and analysis**

Experiment E900a was conducted at the Brookhaven National Laboratory Alternating Gradient Synchrotron (AGS) accelerator. A secondary negative beam ($\pi^-$, $K^-$, $\bar{p}$) of 8.0 GeV/c momentum was tagged with a time-of-flight and Cerenkov-counter identification system. Beams of about 2-4x10$^6$ particles/cycle, with a cycle time of 4.5 seconds and flat top of 2.2 seconds, were incident on a 2x2 cm$^2$ self-supporting 2 mg/cm$^2$ thick $^{197}$Au target. The beam composition was about 98% $\pi^-$, 1% $K^-$, and 1% $\bar{p}$ at the target. The Indiana Silicon Sphere (ISiS) 4$\pi$ detector array [25] was used to measure the light-charged particles and intermediate mass fragments from the reaction. The ISiS detector consists of 162 triple-detector telescopes arranged in a spherical geometry. It covers the polar-angle from 14°- 86.5° in the forward hemisphere and 93.5° - 166° in the backward hemisphere with 74% geometrical acceptance. The detector telescopes consist of a gas-ionization chamber, a 500 μm thick silicon detector, and a 28-mm CsI scintillator with photodiode readout. The gas-ionization chambers provide charge identification for the kinetic energies as low as 1.0 MeV/nucleon. Isotope identification was possible for LCPs (LCPs = H and He) with laboratory kinetic energy per nucleon above $E_k/A \geq 8$ MeV. The analysis involved 2.5x10$^6$ $\pi^-$ and 2.5x10$^3$ $\bar{p}$ events that met the detector array trigger multiplicity requirement ($M_{th} \geq 3$) for thermal particles. Further experimental details can be found in [26-28].

In order to investigate the heating curve for hot nuclei formed in the 8 GeV/c $\pi^-$, $\bar{p}$ + $^{197}$Au reactions, the excitation energy (E*) for the thermal source (defined below) has been calculated on an event-by-event basis according to the prescription,

$$E^* = \sum_{i=1}^{M_{th}} K_i + M_n \langle K_n \rangle + Q + E_g \tag{1}$$

Here $K_i$ is the kinetic energy for each thermal charged-particle in an event of thermal multiplicity $M_{th}$. $M_n$ is the thermal-like neutron multiplicity, which was estimated from the neutron-charged particle correlations reported for 1.2 GeV $\bar{p}$ + $^{197}$Au reactions measured by Goldenbaum et al. [29]. The mean neutron kinetic energy is taken from the correlation between $\langle K_n \rangle$ and $E^*/A$ and then Eq. 1 is iterated to obtain self-consistency. Q is the binding energy difference in the reconstructed event and $E_\gamma$ is a small term to correct for energy released in photons, which is assumed to be $E_\gamma = M_{n(Z \geq 3)} \times 1$ MeV. More details on the calculation of the excitation energy can be found in refs [26,30,31].

The thermal particles used in the calculation of the excitation energy are separated from the fast-cascade/pre-equilibrium contributions based on their kinetic energy in the average source frame, as determined from a two-component moving-source fit to the data [12,13]. The selection of thermal particles, $E_{th}$, makes use of the analysis of data from the 1.8-4.8 GeV $^3$He + $^{197}$Au reaction [12,13]:

$$E_{th} \leq 30 \text{ MeV for } Z = 1 \tag{2}$$

$$E_{th} \leq 9.0Z + 40 \text{ MeV for } Z \geq 2 \tag{3}$$

From this selection, only the thermal-like ejectiles are included in the calculation of the excitation energy.

**III. Results**

**a. Constructing the caloric curve**

In Fig. 1 the kinetic energy spectra for hydrogen and helium ions from the 8 GeV/c $\pi^-$ + Au reactions normalized to solid angle and transformed to the source frame are shown at a forward and a backward angle, and at excitation energies of 4 MeV/nucleon and 8 MeV/nucleon. The very low energy thresholds are achieved by using the gas-ionization chamber/silicon detector stage of the telescopes. The discontinuity near 90 MeV in the H spectra is due to detector punch-through effects. For higher energy Z = 1 particles (above 90 MeV), energies were determined from the energy loss in the CsI crystal. This permitted measurement of "grey particles" up to about 350 MeV, assumed to be protons. The importance of fast-cascade/pre-equilibrium emission can best be seen in the forward angle Z = 2 spectra. For kinetic energies above about 60 MeV, the forward angle spectra show a transition to flatter slopes, indicative of pre-equilibrium emission. In the comparison between the higher and lower excitation energies, the higher excitation energy spectra have flatter slopes in the low energy (thermal) portion of the spectra, as would be expected for a system at a higher temperature. This difference can be seen most clearly in the backward angle Z = 2 spectra.

ISiS provides isotope resolution for particles with kinetic energies $E_k/A \geq 8$ MeV, primarily Z = 1-3 isotopes in these experiments. In Figs. 2 and 3 the spectra for LCPs (LCP = H and He isotopes) are shown for both a forward (22°- 33°) and a backward (128° - 147°) angle range at two excitation energies for the emitting source, $E^*/A$ = 4 and 8 MeV. In all cases the slopes of the spectra in the region of interest, $E_k \sim$ 30-50 MeV (see below), become systematically steeper with increasing LCP mass. This behavior is consistent with the EES model [32], which predicts that the fragment emission time increases with the mass of the fragment; i.e. light particles are emitted early from hotter sources and therefore have higher kinetic energies on average. Because of the large binding energy difference between $^3$He and $^4$He, this effect manifests itself most clearly in the $^3$He/$^4$He spectra. This dependence of slope on fragment mass makes the isotope ratios sensitive to the range of kinetic energies chosen for the calculation, which in turn influences the isotope-ratio temperature, as will be apparent below.

Figs. 2 and 3 also illustrate the expected behavior as a function of excitation energy, i.e. the spectral slopes are flatter at E*/A = 8 MeV than at E*/A = 4 MeV, implying emission from a hotter source. There is little difference between the forward angle data and that at backward angles. This indicates the minimal impact of fast-cascade/pre-equilibrium processes in the acceptance gates for the isotope-ratio calculations.

The isotope-ratio temperatures, $T_{app}$, corresponding to a given excitation energy per nucleon, were calculated with the double-isotope-ratio thermometer [8]. According to Albergo et al, the temperature for a system in chemical and thermal equilibrium can be extracted from a double-isotope-ratio:

$$T_{app} = \frac{B}{\ln(aR)} \tag{4}$$

where B is the binding-energy parameter, a is a factor that depends on statistical weights of the ground state nuclear spins, and R is the ground state population ratio at freeze-out. Tsang et al. [33] have shown that this method is consistent only when B is greater than about 10 MeV, which in practice means that it is essential to use at least one neutron-deficient isotope. Most current caloric curves use $^3$He for this isotope.

The p/d, d/t and $^3$He/$^4$He ratios are plotted in Fig. 4 as a function of kinetic energy. The p/d ratios initially decrease up to kinetic energy $E_k \leq 25$ MeV, followed by a slight increase above this kinetic energy. The d/t ratios increase slightly with increasing energy. A much stronger increase in slope is observed for the $^3$He/$^4$He ratios consistent with known importance of pre-equilibrium $^3$He emission relative to $^4$He and the EES predictions [32,34,35] discussed above. Thus, the value of the double-isotope-ratios are dominated by the $^3$He/$^4$He ratio and it is clear from Eq. (4) that the derived temperature for a given excitation energy will depend on the kinetic energy acceptance of the calculation; the higher the energy cut, the higher the temperature. Unlike the ISiS data of [13], which were obtained with $^3$He beams, the ratios for the $\pi^-$ and $\bar{p}$ beams show little sensitivity to angle, permitting use of the full statistics for the caloric curves presented here.

Caloric curves for the 8 GeV/c $\pi^-$, $\bar{p}$ + $^{197}$Au reaction were constructed from two double-isotope-ratios ($R_{dt-He} = \frac{d/t}{^3He/^4He}$ and $R_{pd-He} = \frac{p/d}{^3He/^4He}$). The apparent isotopic temperatures are given by

$$T_{app-dt} = \frac{14.29}{\ln(1.59 R_{dt-He})} \text{, and} \qquad (5)$$

$$T_{app-pd} = \frac{18.4}{\ln(5.50 R_{pd-He})} \qquad (6)$$

The kinetic-energy acceptance for R was calculated using $^3$He and $^4$He energies between 38-48 MeV. The lower limit was determined by the energy threshold for isotopic resolution of $^4$He for the ISiS telescopes, set by the energy required to punch through the 500μm silicon detector. Accounting for the measured Coulomb shifts between Z = 1 and 2, protons, deuterons, and tritons with energies between 30-40 MeV were taken for the H isotopes in the temperature calculation. The 8 MeV Coulomb shift between Z=1 and Z=2 is imposed as indicated by the location of the evaporative peaks. The upper limits were based on the balance between statistics and the minimization of pre-equilibrium effects. As can be seen from Fig. 5 and Eqs. (5) and (6), because of these high energy cuts, $T_{app}$ values will be higher than for studies that employ lower energy H and He ions.

The upper panel of Fig. 5 shows the $T_{app}$ heating curves for the 8 GeV/c $\pi^-$ + $^{197}$Au reaction from the two thermometers, p/d-$^3$He/$^4$He, and d/t-$^3$He/$^4$He. There is difference in temperature between the two isotopic thermometers at low excitation energy per nucleon that grows in magnitude as the excitation energy increases. Since the $^3$He/$^4$He ratio is the same for both thermometers, the deviation for $E^*/A > 7$ MeV reflects the fact that in this excitation energy region the p/d ratio increases more rapidly than the d/t ratio, as shown in Fig. 4.

Measured yield ratios differ from the primary yield ratios due to sequential decay of the excited fragments. The proton and $^4$He yields, both of which appear in the numerator of R, are enhanced by evaporation from heavier fragments. Tsang et al. [33] have proposed to account for these effects by defining a correction factor $\kappa$ for each isotope ratio by the relationship $R_{app} = \kappa R_0$, where $R_{app}$ and $R_0$ are the measured and equilibrium values of the double-isotope-ratio, respectively. From the definition of equation (4), the temperature of the fragments at freeze-out ($T_0$) can be calculated from:

$$\frac{1}{T_{app}} = \frac{1}{T_0} + \frac{\ln \kappa}{B} \tag{7}$$

Using Tsang's correction factors of $\ln\kappa = 0.0497$, and $0.0097$ for p/d-$^3$He/$^4$He and d/t-$^3$He/$^4$He respectively, the caloric curves from the two thermometers are in much better accord at low excitation energy per nucleon ( $2 \lesssim E^*/A \lesssim 7$ ), as shown in the bottom panel of Fig. 5. However, at higher excitation energies the two thermometers still diverge significantly, a possible consequence of the inability to distinguish between energetic thermal particles and low energy pre-equilibrium particles at high $E^*/A$.

The extracted freeze-out temperatures ($T_0$) increase gradually in the range $1.5 \lesssim E^*/A \lesssim 4$ MeV, followed by a flattering of the slope in the region approximately $4 \lesssim E^*/A \lesssim 9$ MeV for the d/t-$^3$He/$^4$He case and a more distinct plateau in the range $4 \lesssim E^*/A \lesssim 11$ MeV for the p/d-$^3$He/$^4$He thermometer. At higher excitation energies, the caloric curve from the d/t-$^3$He/$^4$He thermometer indicates a more rapid increase in temperature may occur, as in the original ALADIN data [6,7]. The caloric curve from the p/d-$^3$He/$^4$He thermometer does not show such a marked increase. Lower statistics for events with $E^*/A \gtrsim$ 9-10 MeV lead to uncertainties for this portion of the curves.

From experiment E900a, the final data set contained $2.5 \times 10^3$ antiproton events. The investigation of the heating of nuclear matter with 8 GeV/c $\bar{p}$ and $\pi^-$ beams shows significant enhancement of energy deposition in high multiplicity events for $\bar{p}$ compared to $\pi^-$ [26,27,31], thus producing an excitation energy distribution for $\bar{p}$ that extended to higher values than for $\pi^-$. The upper panel of Fig. 6 shows the comparison of the caloric curves derived from the $\pi^-$ data and $\bar{p}$ data using the d/t-$^3$He/$^4$He thermometer. It appears that the caloric curve derived from the $\bar{p}$ data are suggestive of a flatter plateau, although above $E^*/A \gtrsim 3$ MeV, the two curves are statistically the same. However, because of the multiplicity-three trigger level, events with excitation energies less than $E^*/A = 2$ MeV have a greater uncertainty.

A number of observables have been investigated to account for possible differences between the $\pi^-$ and $\bar{p}$ caloric curves. The residue mass of the thermal source from the experiment is well established, especially at $E^*/A \gtrsim 5$ MeV [31]. The lower panel of Fig. 6 shows the mass of the residue nucleus ($A_{res}$) as a function of the excitation energy per nucleon for both $\pi^-$ and $\bar{p}$ data. The reconstructed masses of the residue nuclei decrease linearly with excitation energy for both $\pi^-$ and $\bar{p}$ data and agree within 4% for $\pi^-$ and $\bar{p}$ events. The normalized $Z = 1$ and $Z = 2$ energy spectra are also very similar for both projectiles, except that spectra from the $\bar{p}$ events have lower statistics.

**b. Interpretation**

The dependence on the energy acceptance (energy gates) for selecting particles for the double-isotope ratio temperature calculation is shown in Fig. 7. Here caloric curves from the E900a data are compared using four different energy gates. It can be clearly seen that the higher kinetic energies gates yield higher temperatures from the double-isotope-ratio. This behavior is a direct consequence of the increase in the $^3$He/$^4$He ratio as a function of He energy (Fig. 4), which is the dominant ratio in determining the temperature. Similar behavior for the $^3$He/$^4$He ratio is observed in lower energy hadron-induced reactions [36,37]. The shapes of the caloric curves from the three lowest energy gates (the open circles, the solid triangles and the open triangles) are similar. All three caloric curves show an increase in temperature in the interval $1 \lesssim E^*/A \lesssim 5$ MeV followed by a slope change at higher excitation energies. The slope change which indicates a phase transition disappears for the highest energy gate. This comparison illustrates the sensitivity of temperatures from double-isotope-ratio thermometer to energy acceptance of each isotope pair, and emphasizes that the $T_0$ values determined in this work would be lower if isotopic resolution in ISiS could be obtained at lower kinetic energies.

From the systematic trends shown in Fig. 7 it is estimated that the isotope ratio temperatures would be about 1.5-2.5 MeV lower if extrapolated to the evaporative H and He peaks (Fig. 1). The sequence of caloric curves in Fig. 7 can be interpreted as evidence for 'cooling' of the hot residues as they evolve from the fast cascade stage of the collision toward equilibrium [13, 38]. In this pre-equilibrium/coalescence regime, early emission favors the production of more energetic particles and simpler clusters (e.g. $^3$He relative to $^4$He as in Fig. 4). The net effect is to provide time-dependent snapshots of the 'cooling curve' for the system. Alternatively, this cooling behavior can be described within the framework of the statistical EES model [35], since each particle spectrum consists of a convolution of Maxwellian distributions and the hottest sources are the most significant contributions to the high energy tails of the spectra.

Fig. 8 compares the caloric curves from various experiments with a $^{197}$Au target, which use the double-isotope-ratio based on $^3$He to extract the temperature and the calorimetry method to calculate the excitation energy. In addition to the present data, the plots include data from the ISiS E228 experiment (4.8 GeV $^3$He + $^{197}$Au) [12,13], the EoS collaboration ( 1 GeV $^{197}$Au + C) [9-11], and the ALADIN group (600 MeV/A Au+Au) [6,7]. The curves all fit into a broad band in the temperature-excitation-energy plane. If the cooling curve of Fig. 7 is extrapolated to the peak energies of the evaporative spectrum, then good agreement is observed with the ALADIN data and the disagreement with the EoS data is reduced.

Part of the difference with EoS may be the methodology used for the excitation energy calculation [30]. To investigate the effect of the selection criteria for thermal particles in the excitation energy calculation, the excitation energies from E900a were re-calculated using the same energy selection for thermal particles as the EoS experiment [9,10]. The EoS kinetic energy acceptance for thermal particles is defined as all particles with energies less than $E_k/A \leq 30$ MeV, which yields higher excitation energies than the ISiS prescription for $E^*$ by about 20% [30]. Fig. 9 shows the caloric curves from E900a using this EoS excitation energy definition compared with the E900a caloric curve in the lower panel of Fig. 6 and the Au + C EoS caloric curve. The new ISiS caloric curve is shifted to higher excitation energy, thus producing a flatter and lower caloric curve than that adopted here (lower panel of Fig. 5). The lower $T_0$ values for the EoS relative to the $E^*$-adjusted ISiS data result in part from the lower kinetic energy acceptance for isotope identification in the EoS experiment.

Recent work by Chomaz et. al. [39] investigated the caloric curves and energy fluctuations in a microcanonical liquid-gas phase transition and found that caloric curves may differ depending on the path followed in the thermodynamical variable plane. The size of the systems, which depends on collision dynamics, may also affect the curve. Natowitz [40] has recently investigated a variety of caloric curves and concluded the measurements are self-consistent when the size of the system is taken into account. Thus many reasons may be responsible for the difference in the caloric curves. Each experiment has different formation mechanisms, fragment kinetic energy acceptance, and particle identification limitations due to instrumentation and analysis procedures. As illustrated for the ISiS data in Fig.7, since the kinetic energy cuts for the calculation of R in Eq. (4) are higher than for the other experiments, the $^3$He contribution, and therefore $T_{app}$, are higher.

**IV. Discussion**

In addition to the caloric curve behavior that has been explored in this and other studies, several other signals expected of a liquid-gas phase transition have been observed. Among these are very short disintegration times [41], negative heat capacity [42], and percolation [43]/Fisher scaling [44] analyses of the IMF multiplicity distributions that indicate critical behavior. If we take the total of these results at face value, then the important question remains: is the phase transition in hot nuclei first-order or continuous.

In Fig. 10, we show the relative emission time scale for this reaction [41] overlaid on the $\pi^-$ + $^{197}$Au caloric curve in the lower frame, along with the charge-distribution power-law parameter tau [41] in the upper frame. These results should be viewed in the context of Fisher scaling [44] and percolation analyses [43], both of which show evidence for critical behavior near E*/A ~ 4-5 MeV. On the basis of the angle-integrated d/t-$^3$He/$^4$He caloric curve one would conclude that the phase transition is continuous, although distinct slope changes occur for 4 ≲ E*/A ≲ 9 MeV. From the bottom panel of Fig. 10, the beginning of this slope change region occurs at just the point where the emission time scale becomes minimum, the probability for large cluster size reaches a maximum, and critical behavior is observed. Of course critical behavior and a first order phase transition are mutually exclusive, at least in infinite systems. However, another possible interpretation of Fig. 10 is that once the system reaches the critical energy, its density continues to decrease with $E^*/A$, keeping T constant and implying that the system remains in the vicinity of the critical point over a significant range of excitation energy.

It is also instructive to compare these caloric curve data with nuclear models that assume a phase transition. In Fig. 11, the caloric curves from E900a are compared to predictions from the statistical multifragmentation model (SMM) [45], using the d/t-$^3$He/$^4$He thermometer. The experimentally determined residue charges and masses corresponding to the excitation energy were fed as input to the SMM code and gates were imposed on the calculated spectra to coincide with the experimental acceptance. Fig. 11 shows the experimentally derived apparent temperatures compared with the SMM prediction. The solid square symbols represent the prediction when running the 'cold' version of the code in which the breakup of the system leaves the emitted fragment in a low state of internal excitation, thus suppressing secondary decay. Here the temperatures were calculated with the d/t-$^{3/4}$He ratio using the same energy acceptance as in the experiment, i.e. H at 30-40 MeV and He at 38-48 MeV. For the 'cold' mode, the correction factor for secondary decay is not applied. To see the effect of the energy acceptance, the caloric curve is also derived from the double isotope ratio temperatures by taking all particles that have kinetic energies greater than 5 MeV, represented by open squares. This caloric curve shows more plateau-like behavior for $3 \lesssim E^*/A \lesssim 6$ MeV and has lower temperatures at higher $E^*/A$ than the caloric curve that impose the same energy acceptance as the experimental data.

For comparison, the SMM code was also run in the 'hot' mode, for which some of the available excitation energy resides in the fragments and these pre-fragments can undergo secondary decay. This has the effect of enhancing proton and $^4$He yields and lowering $T_{app}$. The solid triangles are the caloric curve prediction from the 'hot' version of the code. The open triangle represents the caloric curve of the 'hot' mode of SMM in which the correction factor for secondary decay was applied in the double-isotope-ratio calculation. It can be clearly seen that the magnitude and shape of 'cold' version of SMM are in better agreement with the experimental result than the 'hot' version, especially when one accounts for the 'cooling effect' discussed earlier, the data are consistent with the cold version of SMM. The microcanonical temperatures in the SMM calculation are shown by star symbols. The caloric curve from microcanonical temperatures is in better agreement with 'cold' mode than 'hot' mode. This implies that isotopic temperatures are strongly affected by the secondary decays of hot primary fragments.

However, the SMM does not take into account the light charged particles which could be emitted during the thermal expansion of the nuclear system before the fragment formation [46]. Therefore, the final model analysis should involve a comparison of the energy spectra of the emitted particles.

## V. Summary

In summary, hadron-induced reactions provide transparent systems for investigating the thermal properties of highly excited nuclei, with minimal contributions from dynamical effects such as compression and rotation. This paper presents caloric curves from the 8 GeV/c $\pi^-$ and $\bar{p}$ + $^{197}$Au reactions. The caloric curves have been derived from reconstructed excitation energies and two double-isotope-ratio temperatures, d/t-$^3$He/$^4$He and p/d-$^3$He/$^4$He. Contributions to the temperatures from secondary decay have been accounted for with empirical correction factors [33]. The caloric curves from the two thermometers (p/d-$^3$He/$^4$He and d/t-$^3$He/$^4$He) overlap below E*/A ~ 7 MeV when correction factors are applied, but diverge significantly at higher excitation energy. The lack of ability to converge the two temperatures at the higher excitation energies indicates that more work is needed to understand the effects of secondary decay for the higher excitation energies.

The freeze-out temperatures ($T_0$) increase gradually in the temperature range 1.5 ≲ E*/A ≲ 5 MeV, followed by a slope change at approximately 4 ≲ E*/A ≲ 9 MeV, where the emission time scale becomes very short, IMF multiplicities increase rapidly and extra expansion energy is observed. An increase in temperature at higher excitation energy is also observed suggesting evidence for a phase transition. The double-isotope-ratio temperature depends strongly on the kinetic energies of the particles that are included in the isotope ratio calculation. As higher energy particles are included, higher temperatures result from the calculation, in essence, providing a time-dependent cooling curve for the hot residues as they evolve toward equilibrium. Although the overall temperatures depend on this choice of energy acceptance the basic character of the caloric curve remains unchanged.

Besides the kinetic energy acceptances imposed on the spectra, the definition of thermal particles in the E* calculation can result in different caloric curves. Since each measurement imposes a separate set of acceptances, the caloric curves from different experiments do not totally overlap with one another. However, they all fit in to a broad band on the temperature-excitation-energy plane. When comparing with other data one must take care to take into account any differences in energy acceptance of both the particles included in the calculation of the temperature and those included in the excitation energy calculation. The effect of these experimental parameters is significant as has been demonstrated by careful analysis of the current data.

Overall the data from E900 are high quality data which represent the highest excitation energy data for large systems (A>100) [40]. The data are minimally affected by compression, angular momentum or flow and reach high excitation energies with moderate changes in system size. These attributes of the data make the caloric curves presented in this paper an important addition to the existing body of data. While the initial rise and plateau have been seen in many systems, the second rise above $E^*/A \sim 9$ MeV is of significant interest. This rise is consistent with a phase transition over a narrow range in $E^*/A$ and is nicely complemented by the Fisher Droplet analysis [44] and the percolation analysis [43], which also find a phase transition and evidence for critical behavior in this region.

**VI. Acknowledgements**

The authors wish to thank R.N. Yoder, W. Lozowski, J. Vanderwerp and T. Hall of Indiana University; and P. Pile, J. Scaduto, L. Toler, J. Bunce, J. Gould, R. Hackenberg, C. Woody, F. Kobasiuk and T. Mruczkowski from AGS for their assistance with the experiments. This work was supported by the U.S. Department of Energy, the National Science Foundataion, the National and Engineering Research Council of Canada, Grant No. P03B 048 15 of the Polish State Committee for Scientific Research, Indiana University Office of Research and the University Graduate School, Simon Fraser University and the Robert A. Welch Foundation.

Fig. 1: Forward- and backward-angle kinetic energy spectra in the source frame for Z = 1 (top panels) and Z = 2 (bottom panels) at E*/A = 4 MeV (left panels) and E*/A = 8 MeV (right panels). Statistical uncertainties are shown as error bars but most of them are smaller than the symbols.

Fig. 2: Forward- and backward-angle kinetic energy spectra for p, d, and t at E*/A = 4 MeV (left panels) and E*/A = 8 MeV (right panels). Statistical uncertainties are shown as error bars but most of them are smaller than the symbols.

Fig. 3: Forward- and backward-angle kinetic energy spectra for $^3$He and $^4$He at E*/A = 4 MeV (left panels) and E*/A = 8 MeV (right panels). Statistical uncertainties are shown as error bars but most of them are smaller than the symbols.

Fig. 4: p/d, d/t and $^3$He/$^4$He isotope yield ratios as a function of kinetic energy in the average source frame for forward-angle (open circles) and backward-angle (solid circles) at E*/A = 4 MeV (top panels) and E*/A = 8 MeV (bottom panels).

Fig. 5: Caloric curve for 8 GeV/c $\pi^-$+$^{197}$Au from p/d-$^3$He/$^4$He and d/t-$^3$He/$^4$He thermometers using apparent yield to calculate temperature (top panel) and temperature corrected for secondary decay (bottom panel).

Fig. 6: Top panel shows the comparison of caloric curves using d/t-$^3$He/$^4$He thermometer corrected for secondary decay from 8 GeV/c $\pi^-$+$^{197}$Au (solid circles) and 8 GeV/c $\bar{p}$ +$^{197}$Au (open circles) reactions. Bottom panel compares the residue mass of pion data (crosses) with antiproton data (open squares). Statistical uncertainties are shown as error bars but most of them are smaller than the symbols.

Fig. 7: The caloric curve for 8 GeV/c $\pi^-$+$^{197}$Au from the d/t-$^3$He/$^4$He thermometer corrected for secondary decay using four different kinetic energy acceptances as given in the graph in MeV.

Fig. 8: Caloric curves from the E900a experiment (8 GeV/c $\pi^-+^{197}$Au, same data as in bottom panel of Fig. 5), ISiS E228 experiment (ref 12, 13), ALADIN data (ref 6,7), and EoS data (ref 9-11).

Fig. 9: Caloric curve from the 8 GeV/c $\pi^-+^{197}$Au reaction using the d/t-$^3$He/$^4$He thermometer corrected for secondary decay (solid circles); using the EoS excitation energy calculation (open circles); and from EoS collaboration experiment (open triangles).

Fig. 10: The top panel shows the charge-distribution power-law parameter as a function of excitation energy per nucleon from ref [41]. The bottom panel shows the caloric curve using the d/t-$^3$He/$^4$He thermometer, corrected for secondary decay from the 8 GeV/c $\pi^-+^{197}$Au reaction overlaid on the relative emission time scale for the reaction from ref [41].

Fig. 11: The caloric curve using the d/t-$^3$He/$^4$He thermometer, corrected for secondary decay from the 8 GeV/c $\pi^-+^{197}$Au reaction, (open circles); 'cold' version of SMM prediction using the d/t-$^3$He/$^4$He thermometer and imposed the same energy acceptance as in the experimental data (solid squares); 'cold' SMM version without imposed the energy acceptance (open squares); 'hot' SMM version without the correction factor for secondary decay (solid triangles), and with correction for secondary decay (open triangles); and SMM prediction from the microcanonical temperature (stars).

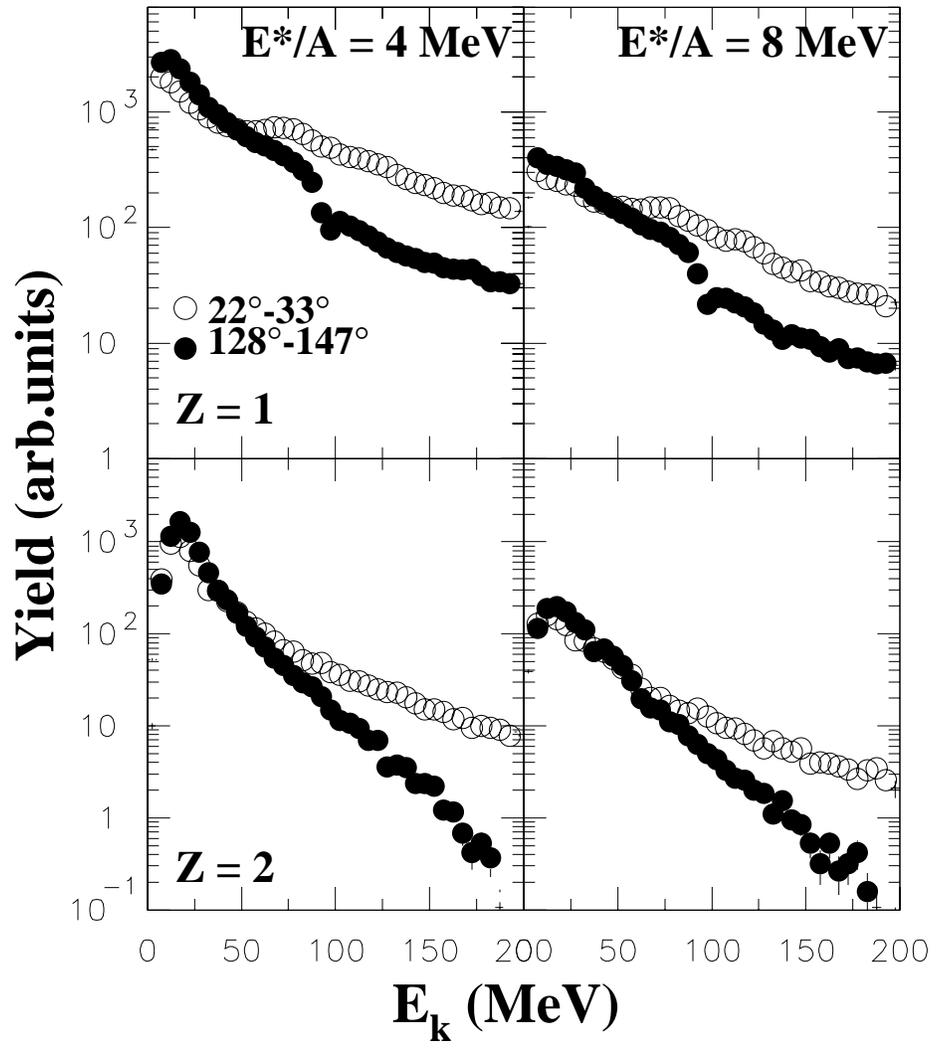

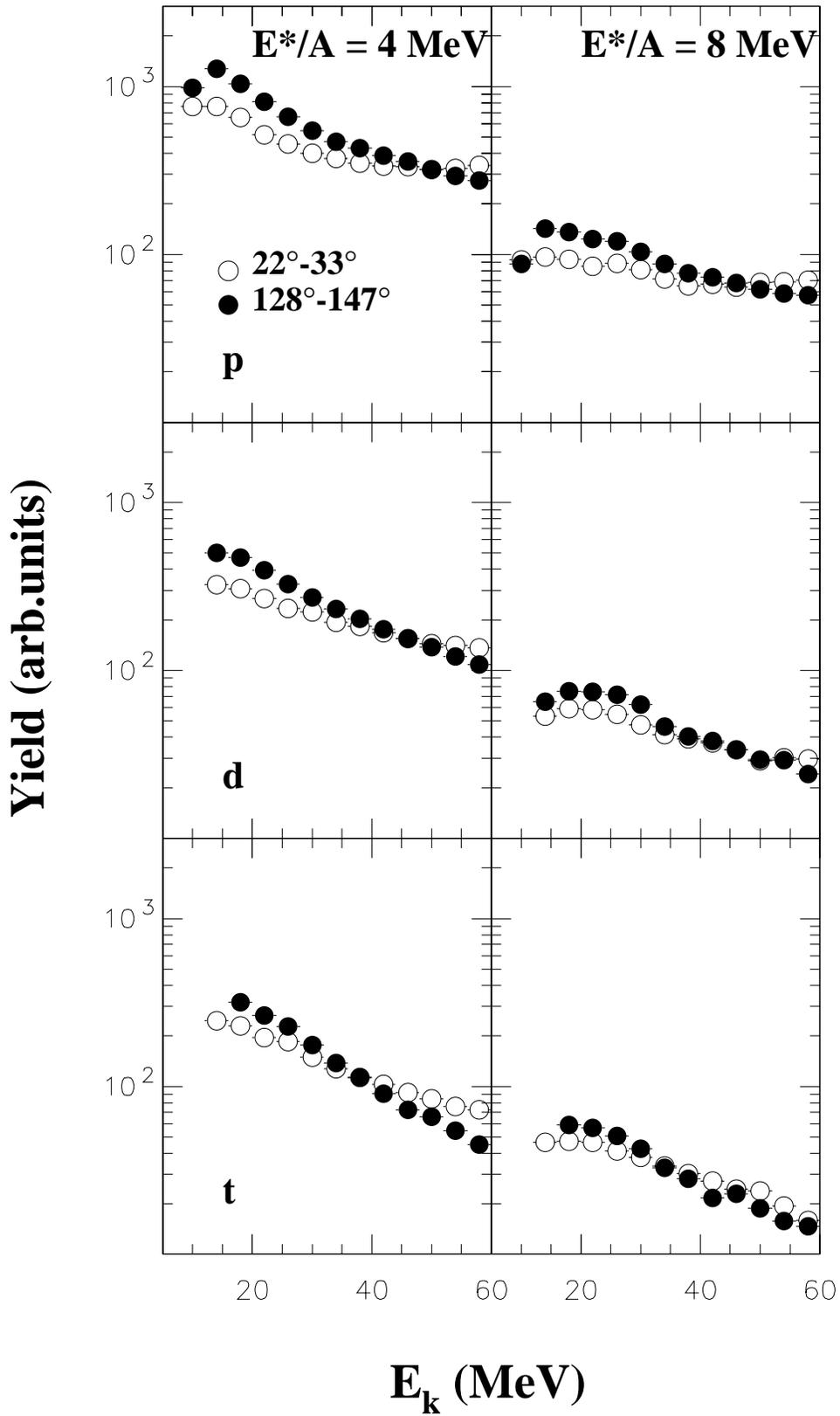

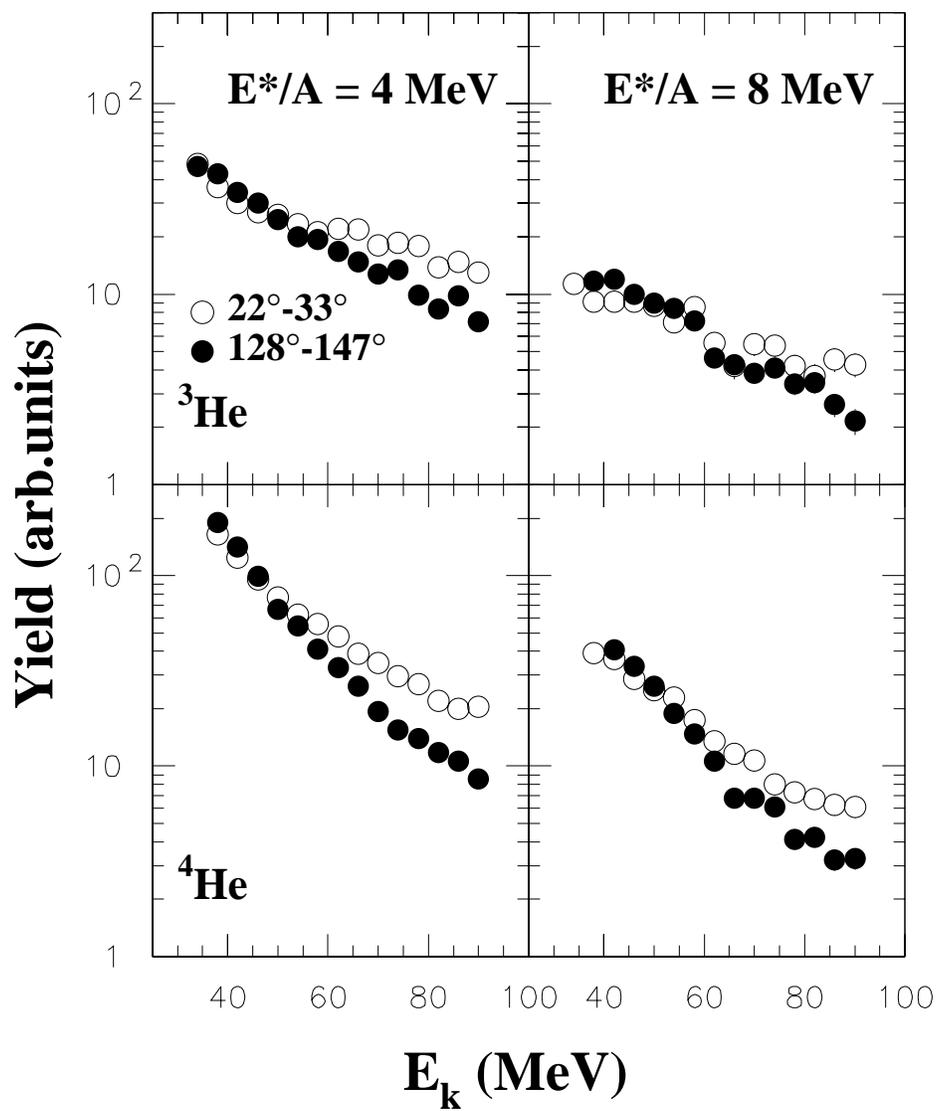

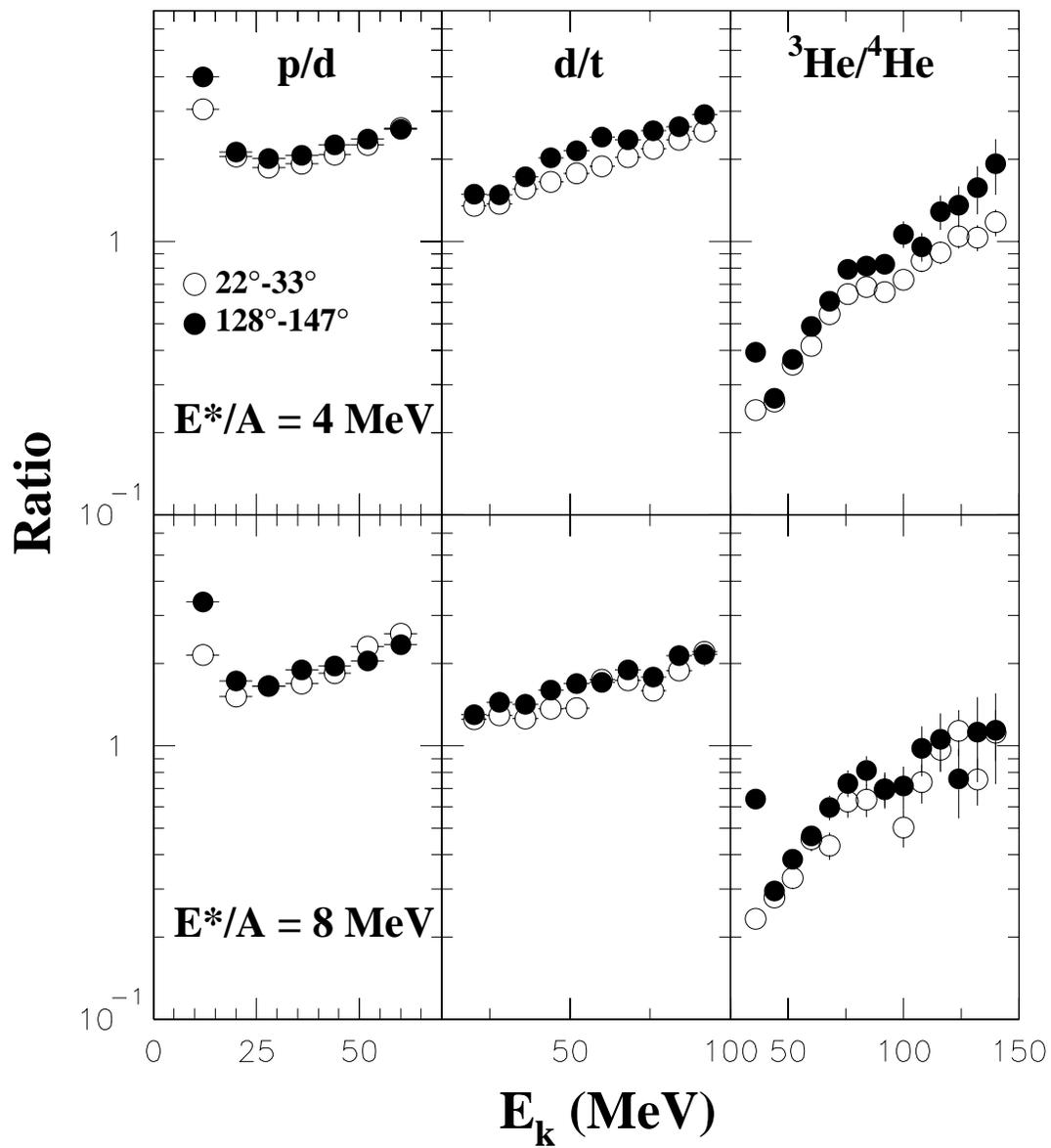

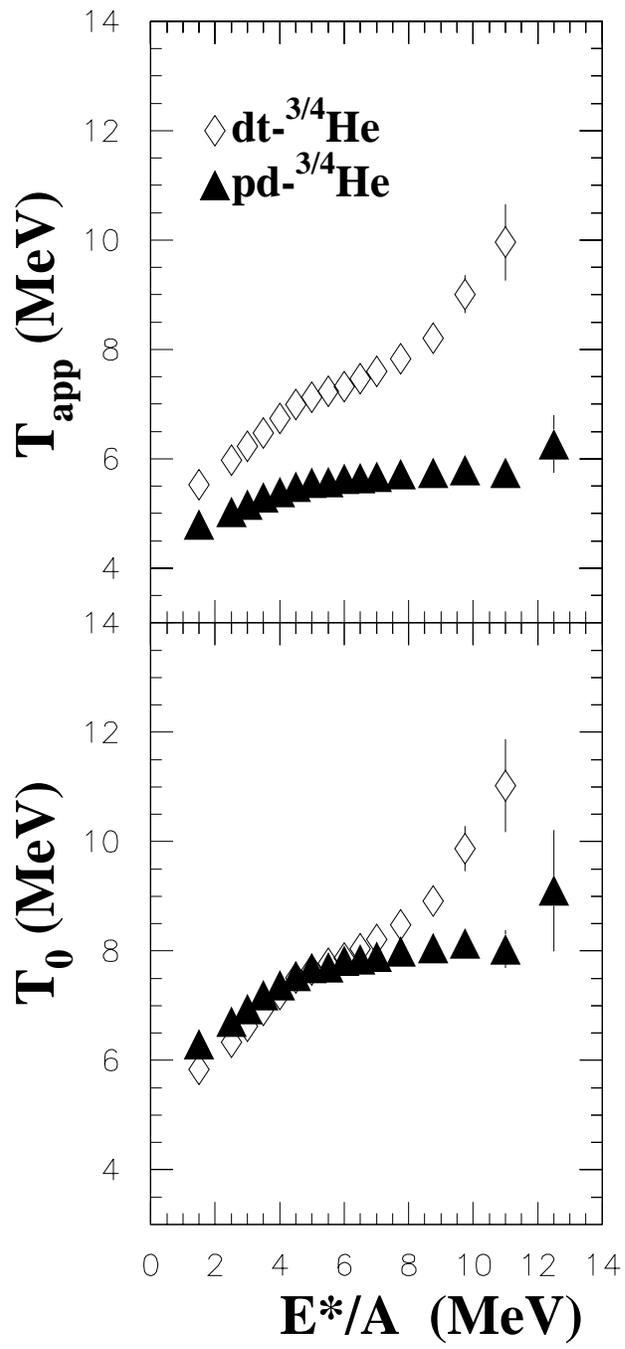

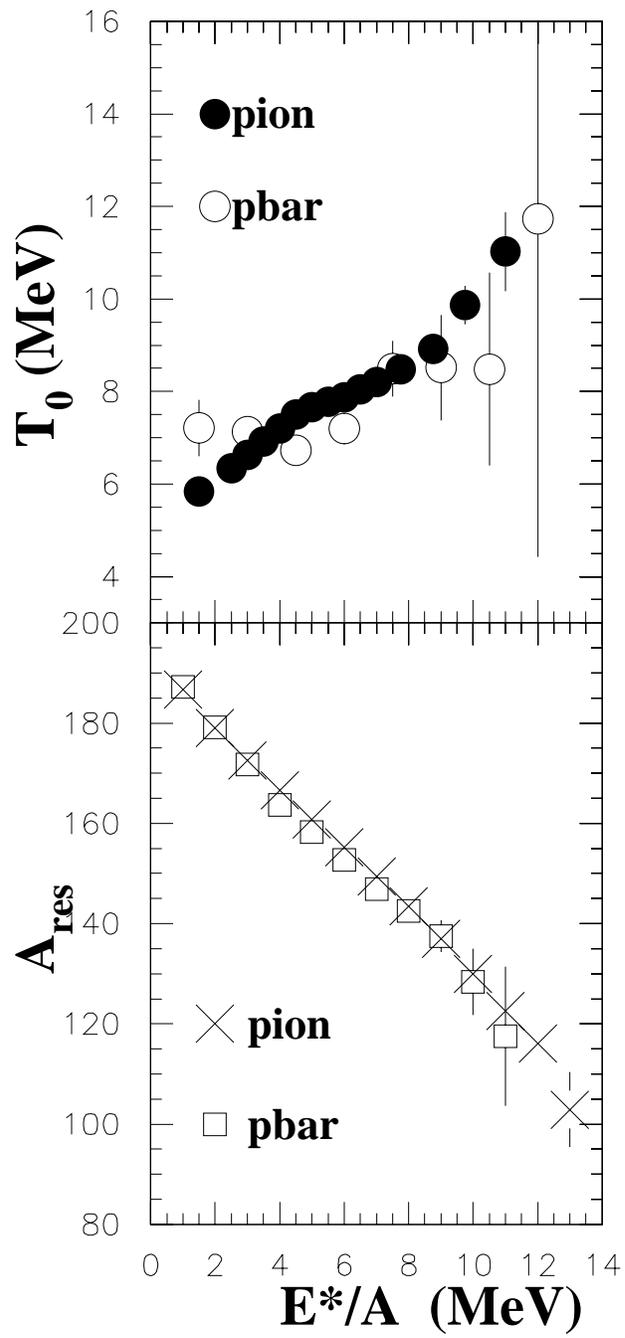

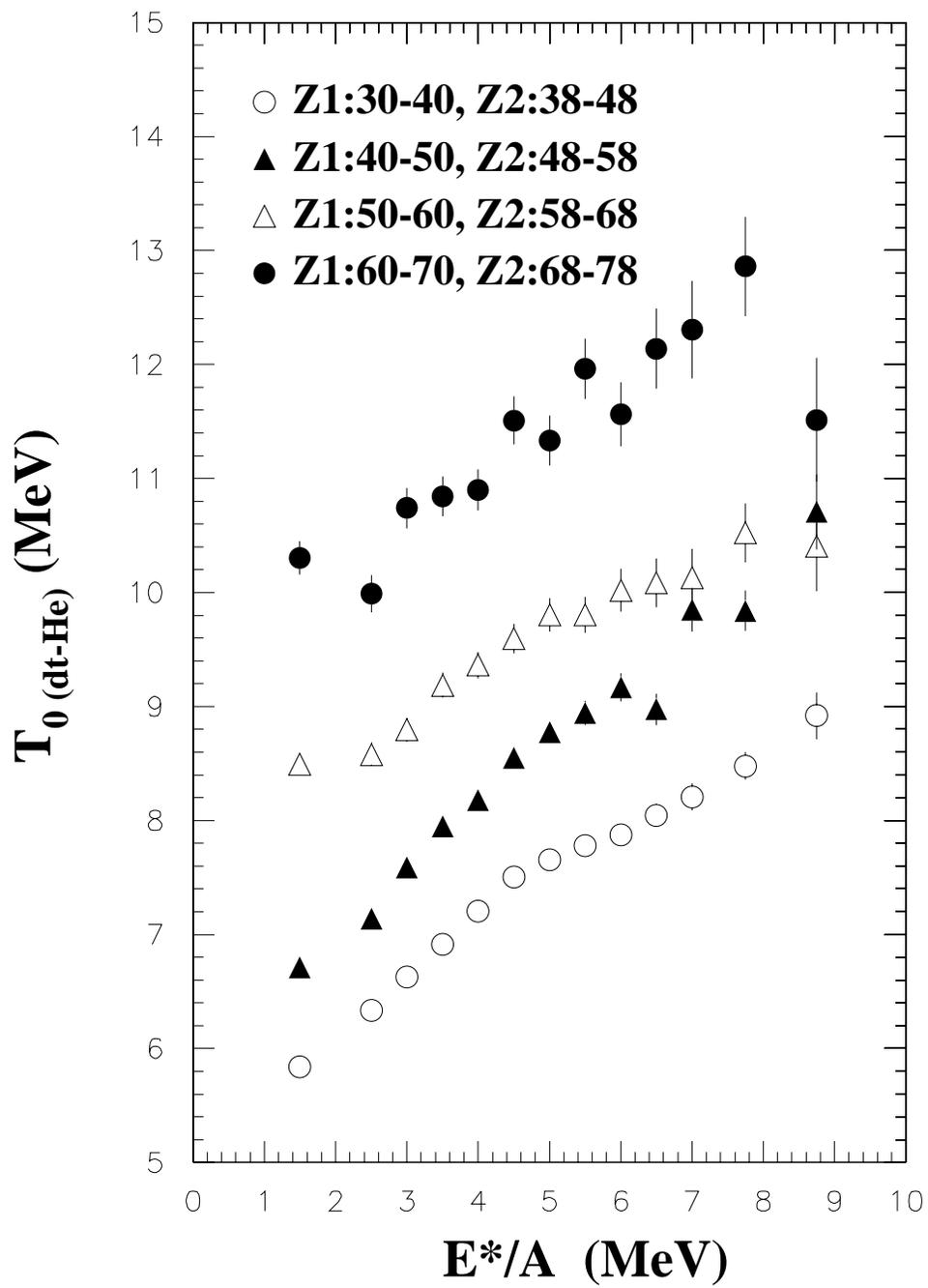

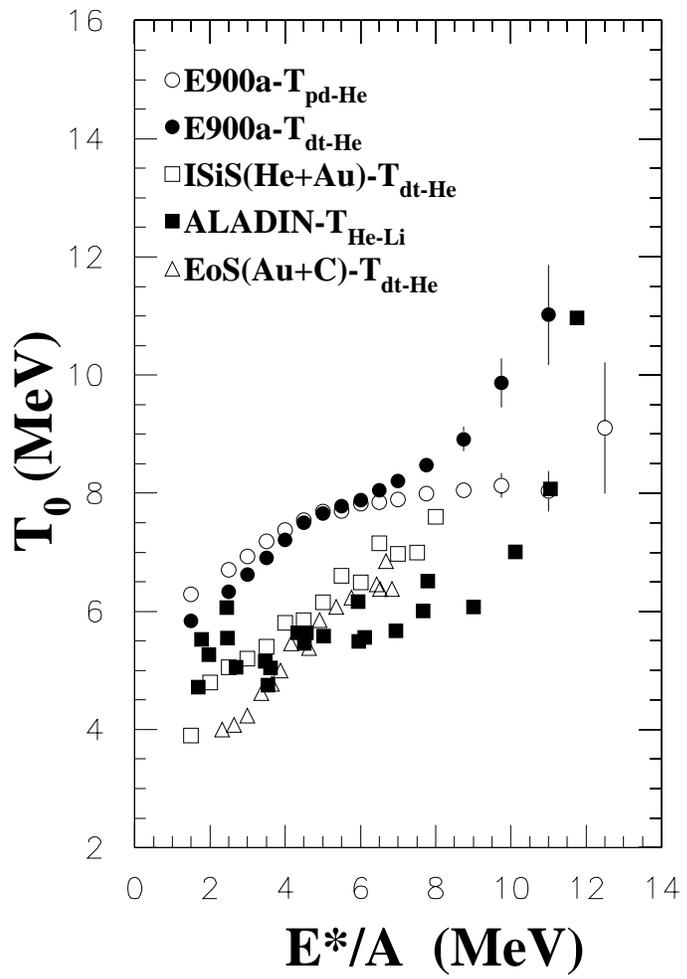

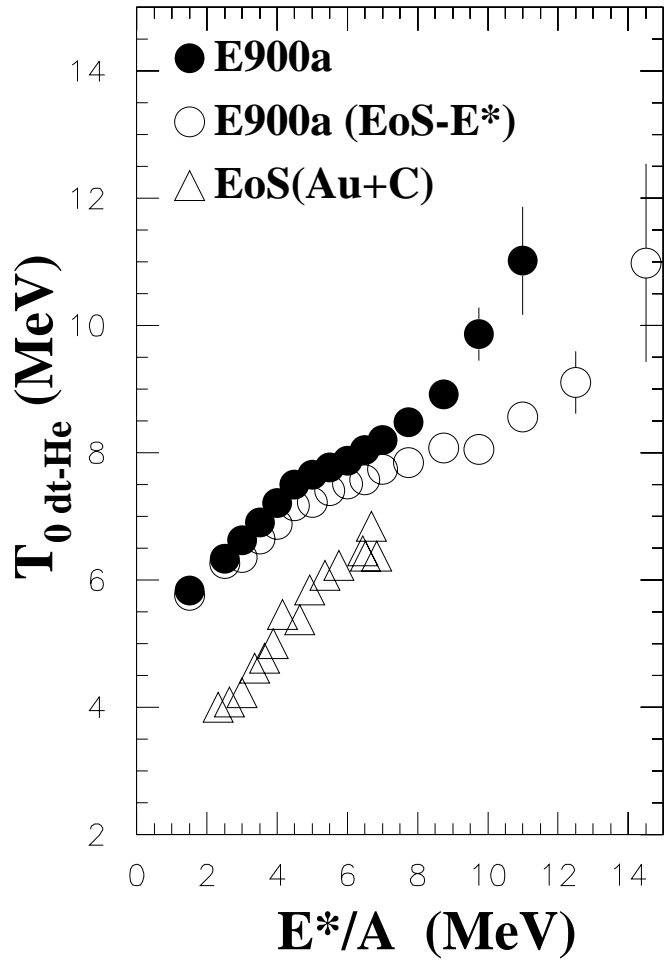

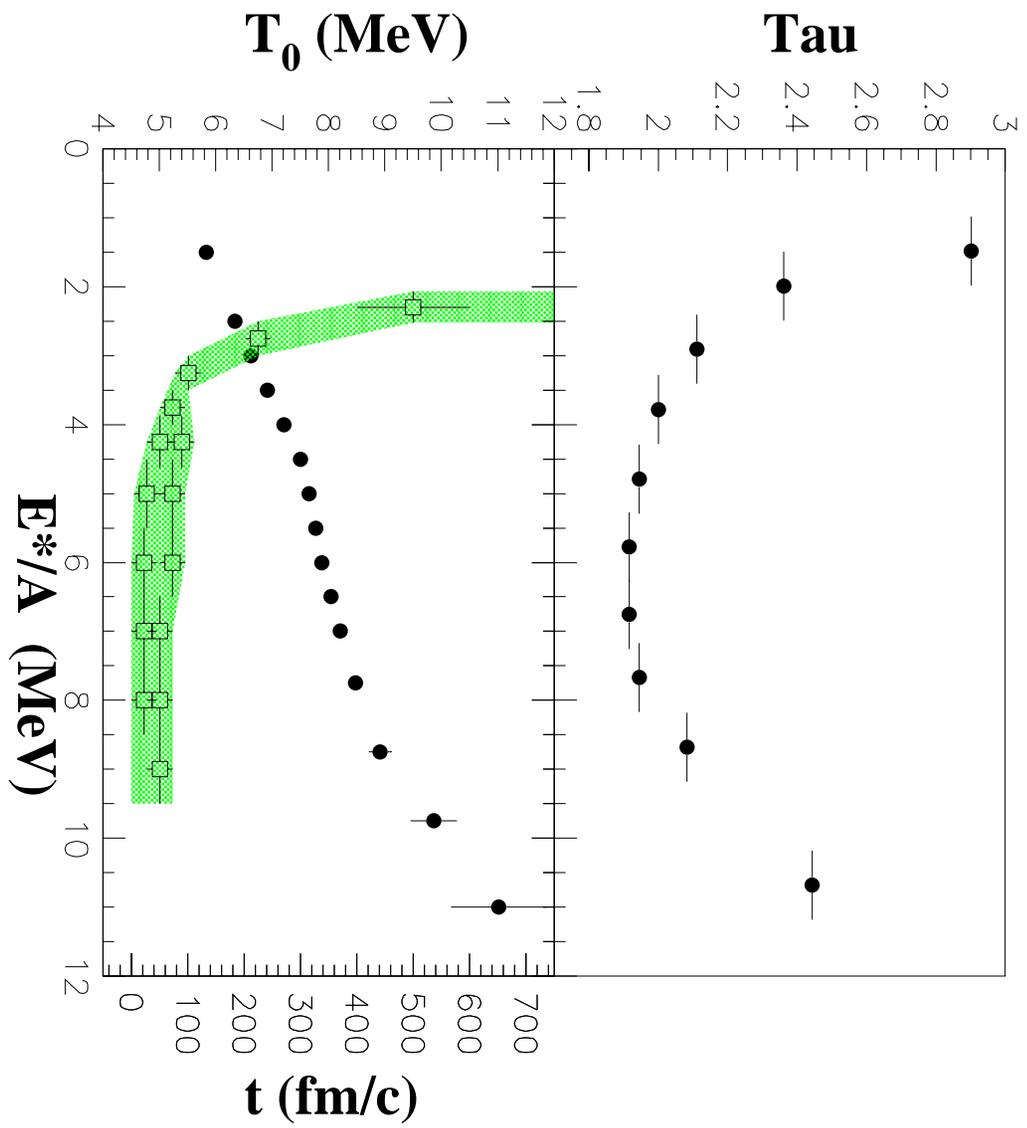

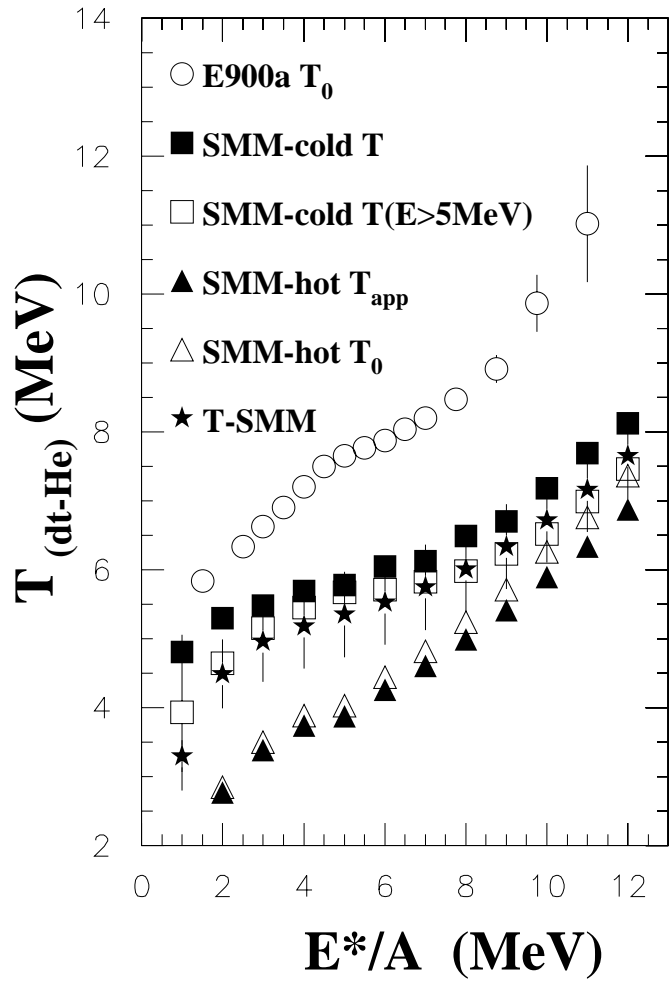